\documentclass{article}
\usepackage{graphicx}
\usepackage{amsmath}
\input{psfig.sty}
\newtheorem{theorem}{Theorem}
\newtheorem{acknowledgement}[theorem]{Acknowledgement}

\input{tcilatex}

\begin{document}

\title{Evolutionary Dynamics of the World Wide Web}
\author{Bernardo A. Huberman and Lada A. Adamic \\
%EndAName
Xerox Palo Alto Research Center\\
Palo Alto, CA 94304}
\maketitle

\begin{abstract}
We present a theory for the growth dynamics of the World Wide Web that takes
into account the wide range of stochastic growth rates in the number of
pages per site, as well as the fact that new sites are created at different
times. This leads to the prediction of a universal power law in the
distribution of the number of pages per site which we confirm experimentally
by analyzing data from large crawls made by the search engines Alexa and
Infoseek. The existence of this power law not only implies the lack of any
length scale for the Web, but also allows one to determine the expected
number of sites of any given size without having to exhaustively crawl the
Web.\pagebreak
\end{abstract}

The World Wide Web (Web) has become in a very short period one of the most
useful sources of information for a large part of the world's population.
Its exponential growth, as shown in Figure 1, from a few sites in 1994 to
millions today, has transformed it into an ecology of knowledge in which
highly diverse information is linked in extremely complex and arbitrary
fashion (1). Moreover, several estimates of the total number of pages (2)
indicate that due to the rapid growth of the Web, most search engines are
only finding a fraction of all the available sites (2).

\bigskip
%% \FRAME{dtbpFU}{4.8577in}{3.2085in}{0pt}{\Qcb{Figure1: The growth of
%% the World Wide Web}}{}{figure1.eps}
%%{\special{language "Scientific Word";type
%%"GRAPHIC";maintain-aspect-ratio TRUE;display "USEDEF";valid_file "F";width
%%4.8577in;height 3.2085in;depth 0pt;original-width 7.1555in;original-height
%%4.7167in;cropleft "0";croptop "1";cropright "1";cropbottom "0";filename
%%'figure1.eps';file-properties "XNPEU";}}
\psfig{file=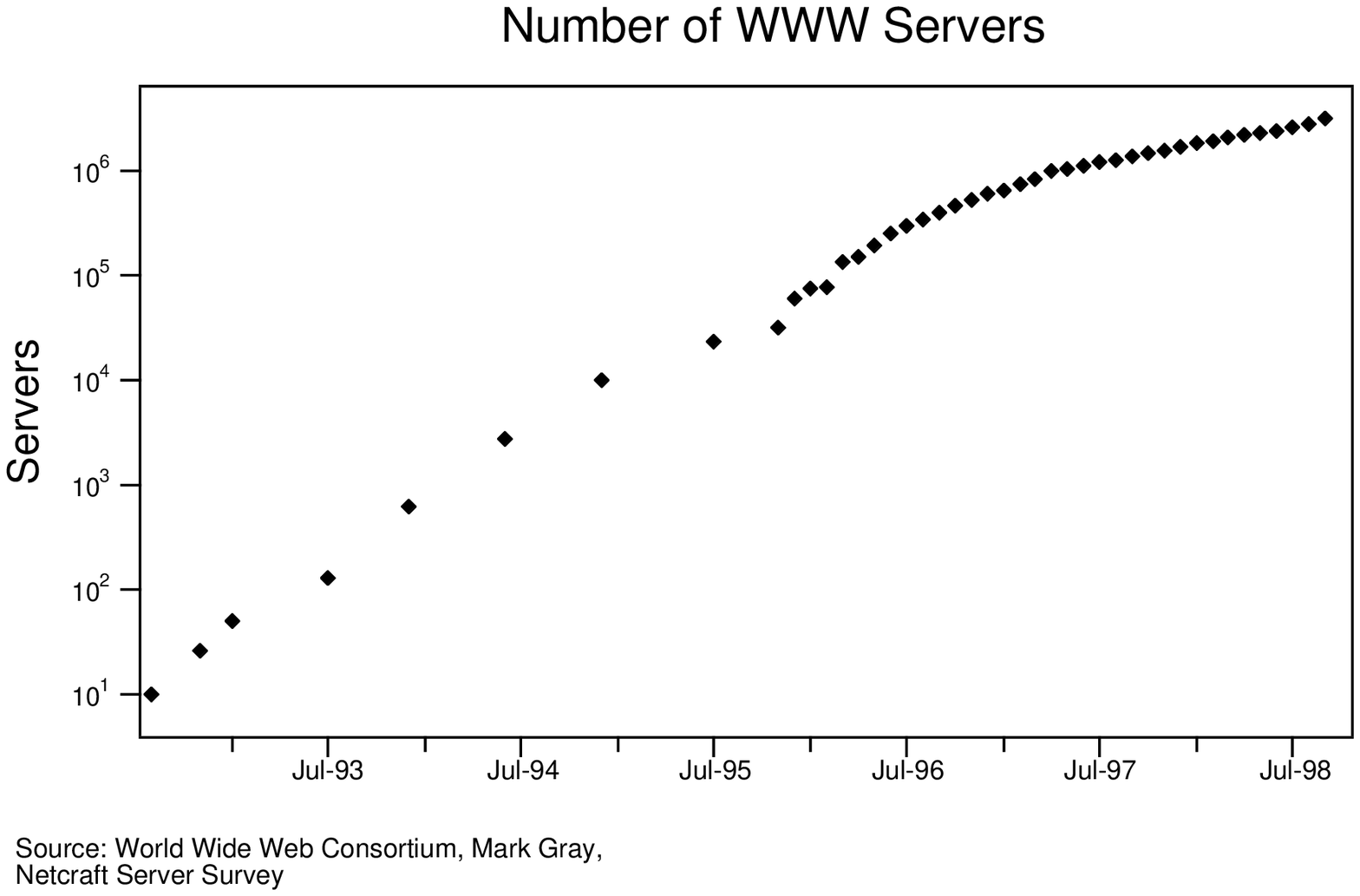,width=4.8577in,height=3.2085in}

In order to develop an evolutionary theory of the growth of the Web, we\
first consider the number of pages belonging to a given site as a function
of time. Since pages within sites are typically organized in hierarchical,
tree-like, fashion, the number of pages added at any given time to a site
will be proportional to those already existing there. Thus, if $n_{s}(t)$ is
the number of pages belonging to a site $s$ at time $t$, the number at the
next interval of time, $n_{s}(t+1)$, is determined by 
\begin{equation}
n_{s}(t+1)=n_{s}(t)+g(t+1)n_{s}(t)  \tag{1}
\end{equation}
where $g(t)$is the growth rate. Given the unpredictable character of site
growth, we assume that $g(t)$ fluctuates in an uncorrelated fashion from one
time interval to the other about a positive mean value $g_{0}$. In other
words 
\begin{equation}
g(t)=g_{0}+\xi (t)  \tag{2}
\end{equation}
with the fluctuations in growth, $\xi (t),$behaving in such a way that $<\xi
(t)>=0$ and $<\xi (t)\xi (t+1)>=2\sigma \delta _{t,t+1}$, i.e. they are
delta correlated and with zero mean. This assumption was confirmed by a
study of the growth of the Xerox Corp. Web site, whose fluctuations in
growth are plotted in Figure 2. Pearson's correlation test accepts at the
0.05 level (with p-value 0.71) the hypothesis that the day to day
fluctuations in the growth rate are uncorrelated.

%%\FRAME{dtbpFU}{4.0378in}{3.4636in}{0pt}{\Qcb{Figure 2: \ Rate of growth of
%%the Web site for the Xerox Corporation.}}{}{figure2.eps}{\special{language
%%"Scientific Word";type "GRAPHIC";display "USEDEF";valid_file "F";width
%%4.0378in;height 3.4636in;depth 0pt;original-width 6.9626in;original-height
%%5.5391in;cropleft "0";croptop "1";cropright "1";cropbottom "0";filename
%%'figure2.eps';file-properties "XNPEU";}}
\psfig{file=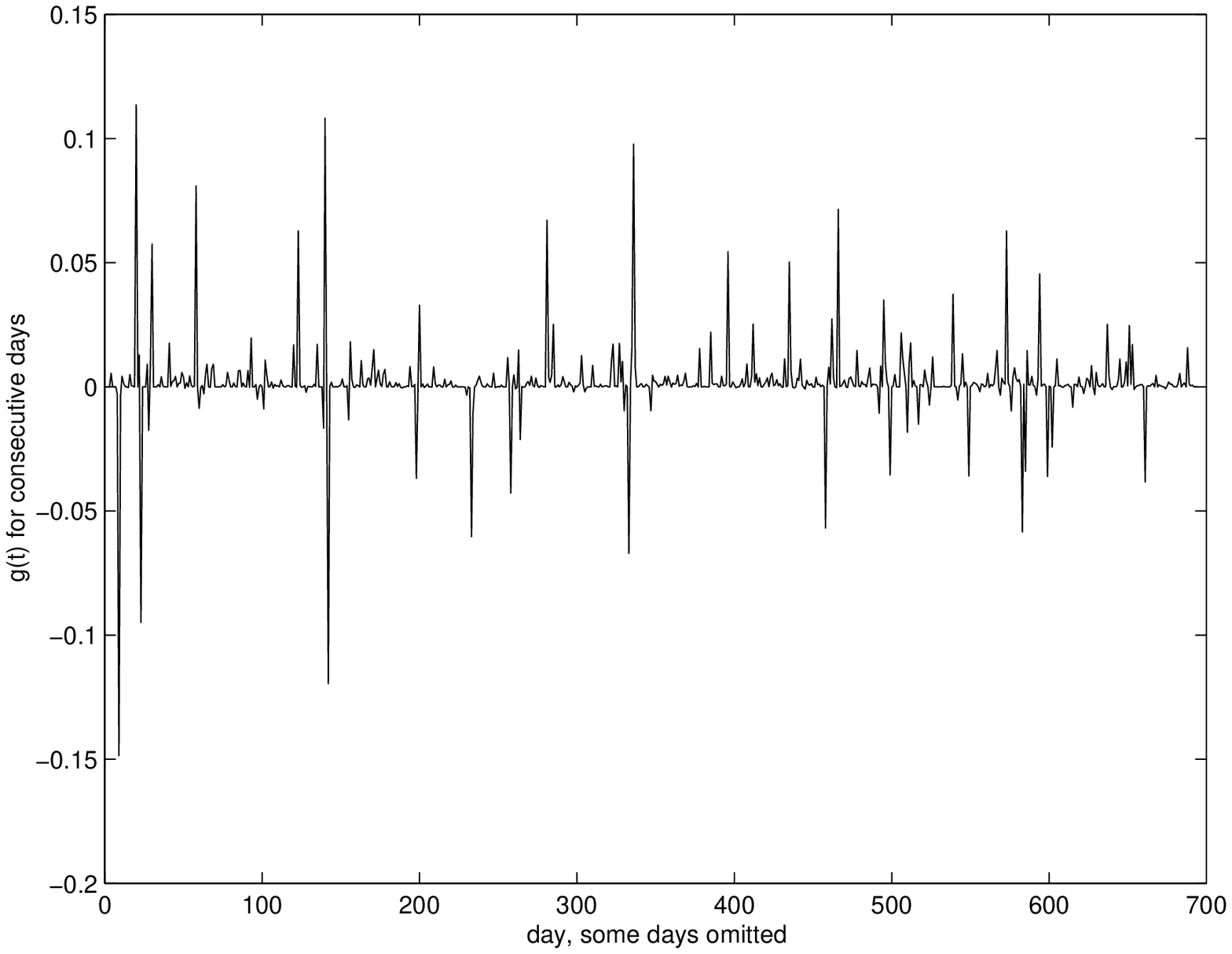,width=4.0378in,height=3.4636in}

In order to obtain the distribution of pages per site, we sum Eq. (1) to get 
\begin{equation}
\sum_{t=0}^{T}\frac{n_{s}(t+1)-n_{s}(t)}{n_{s}(t)}=\sum_{t=0}^{T}g(t) 
\tag{3}
\end{equation}
Changing the sum to an integral (which assumes that the differences in pages
between two time steps is small) we obtain 
\begin{equation}
\int_{0}^{T}\frac{dn_{s}}{n_{s}}=\ln \frac{n(T)}{n_{s}(0)}=\sum_{t=0}^{T}g(t)
\tag{4}
\end{equation}
Notice that the right hand side of Eq. (4) is a sum over discrete time
steps, at each of which we assume the values of g to be normally distributed
with mean $g_{0}$ and variance $\sigma ^{2}$. This corresponds to a Brownian
motion process with stationary and independent increments. By invoking the
Central Limit Theorem we can assert that for every time step $t$, the
logarithm of $n_{s}$ is normally distributed with mean $g_{0}t$ and variance 
$\sigma ^{2}t$(3,4). This means that the distribution of the number of pages
for sites created at the same time and with the same average growth rate is
log-normal (5), i.e, its density is given by 
\begin{equation}
P(n_{s})=\frac{1}{n_{s}\sqrt{t}\sqrt{2\pi \sigma ^{2}}}\exp -[\frac{(\ln
n_{s}-g_{0}t)^{2}}{2\sigma ^{2}t}]  \tag{5}
\end{equation}
where the time dependent drift $g_{0}t$ is the mean of $\ln n_{s}$ ,
reflecting the fact that as time goes on there are more pages added on
average than deleted. The variance of this distribution is related to the
median $m=\exp (g_{0}t)$ by $Var(n_{s})=m^{2}\exp \left( t\sigma ^{2}\right)
(\exp \left( t\sigma ^{2}\right) -1).$

Some insight into the dynamics of this growth can be obtained by noticing
that the stochastic differential equation associated with Eq. (1), which is
given by 
\begin{equation}
\frac{dn_{s}}{dt}=[g_{0}+\xi (t)]n_{s}  \tag{6}
\end{equation}
can be solved exactly(6). The solution is the stochastic growth process 
\begin{equation}
n_{s}(t)=n_{s}(0)\exp \left( g_{0}t+w_{t}\right)  \tag{7}
\end{equation}
where $w_{t}$ is a Wiener process such that $<w_{t}>=0$ and $%
<w_{t}^{2}>=\exp \sigma ^{2}t$. Equation (7) shows that typical fluctuations
in the growth of the number of pages away from their mean rate $g_{0}$ relax
exponentially to zero. On the other hand, the $n^{th}$moments of $n_{s}$ ,
which are related to the probability of very unlikely events, grow in time
as $<n_{s}(t)^{n}>=[n_{s}(0)]^{n}\exp [n(n-\sigma g_{0}t]$\ , indicating
that the evolutionary dynamics of the web is dominated by occasional bursts
in which large number of pages suddenly appear at a given site. These bursts
are responsible for the long tail of the probability distribution and make
average behavior to depart from typical realizations(7).

In order to consider the evolutionary dynamics of the whole Web, it is
important to notice that the distribution of the number of pages depends on
the time that has elapsed since the site was created. Since the number of
sites in the Web has doubled on average every six months, newer sites are
more numerous than older one, and therefore the distribution of pages per
site, for all sites of a given growth rate regardless of age, is a mixture
of lognormals given by Equation (5), whose age parameter $t$ is weighted
exponentially. Thus, in order to obtain the true distribution of pages per
site that grow at the same growth rate, we need to compute the mixture given
by 
\begin{equation}
P(n_{s})=\int \lambda \exp (\lambda t)\frac{1}{n_{s}\sqrt{2\pi t\sigma ^{2}}}%
\exp -[\frac{(\ln n_{s}-g_{0}t)^{2}}{2t\sigma ^{2}}]dt  \tag{8}
\end{equation}
which can be calculated analytically to give 
\begin{equation}
P(n_{s})=Cn_{s}^{-\beta }  \tag{9}
\end{equation}
where the constant $C$ is given by $C=\lambda /\sigma (\sqrt{(g_{0}/\sigma
)^{2}+2\lambda }$ and the exponent $\beta $ is in the range $[-\infty ,-1]$
and determined by $\beta =-1+\frac{g_{0}}{\sigma ^{2}}-\frac{\sqrt{%
g_{0}^{2}+2\lambda \sigma ^{2}}}{\sigma ^{2}}$.

\bigskip

Lastly we need to take into account different growth rates for sites of the
Web, since the distribution given by Eq. (9) applies only to sites that have
the same growth rate $g=g(g_{0},\sigma )$. Since each growth rate occurs
with a particular probability $P(g)$, and gives rise to a power law
distribution in the number of pages per site with a specific exponent, the
probability that a given site with an unknown growth rate has $n_{s}$ pages
is given by the sum, over all growth rates $g$, of the probability that the
site has so many pages given $g$, multiplied by the probability that a
site's growth rate is $g$, i.e. 
\begin{equation}
P(n_{s})=\sum_{i}P(n_{s}|g_{i})P(g_{i})  \tag{10}
\end{equation}
Since we have already shown that each particular growth rate gives rise to a
power law distribution with a specific value of the exponent $\beta (g)$,
this sum is of the form 
\begin{equation}
P(n_{s})=\frac{c_{1}}{n_{s}^{\beta _{1}}}+\frac{c_{2}}{n_{s}^{\beta _{2}}}%
+...+\frac{c_{n}}{n_{s}^{\beta _{n}}}  \tag{11}
\end{equation}
which, for large values of $n_{s}$ behaves like a power law with an exponent
given by the smallest power present in the series.

We thus obtain the very general result that the evolutionary dynamics of the
World Wide Web gives rise to an asymptotic self similar structure in which
there is no natural scale, with the number of pages per site distributed
according to a power law. This implies that on a log-log scale, the number
of pages per site, for large $n$, should fall on a straight line.

In order to test this theory, we studied data generated by crawls of the
World Wide Web made by two search engines, Alexa (8) and Infoseek(9), which
covered 259,794 and 525,882 sites respectively. The plots in Figure 3 show
the probability of drawing at random from the sites in the crawls a site
with a given number of pages. Both data sets display a power law over
several orders of magnitude, with a drop-off at approximately $10^{5}$
pages, which is due to the fact that crawlers don't systematically collect
more pages per site than this bound because of server limitations. The power
law, as well as the drop-off are illustrated in Figure 3.

\begin{equation*}
%%\FRAME{itbpFU}{4.056in}{3.4117in}{0.0104in}{\Qcb{Figure 3: \ Distribution of
%%pages per site.}}{}{figure3.eps}{\special{language "Scientific Word";type
%%"GRAPHIC";maintain-aspect-ratio TRUE;display "USEDEF";valid_file "F";width
%%4.056in;height 3.4117in;depth 0.0104in;original-width
%%6.9047in;original-height 5.802in;cropleft "0";croptop "1";cropright
%%"1";cropbottom "0";filename 'figure3.eps';file-properties "XNPEU";}}
\psfig{file=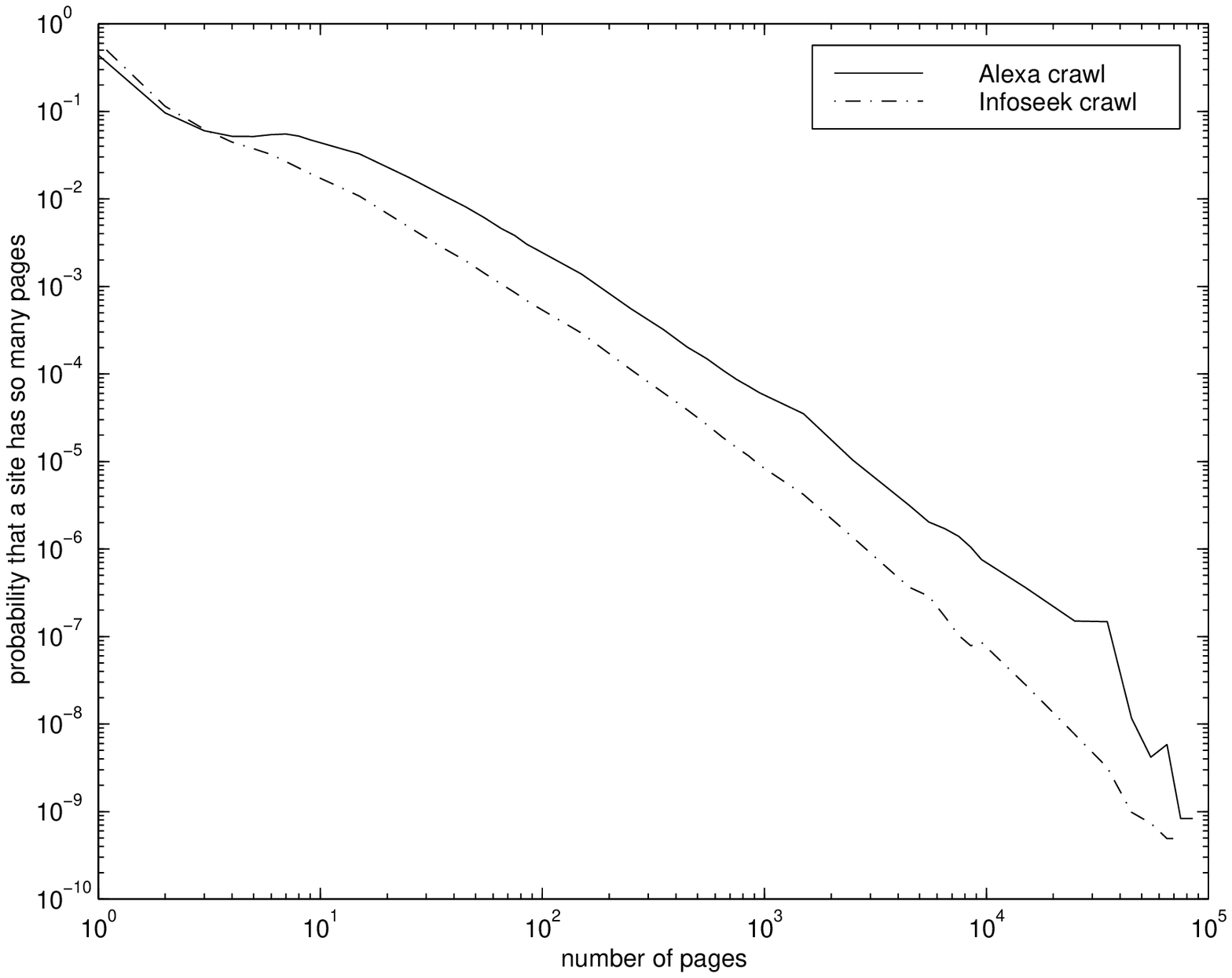,width=4.056in,height=3.4117in}
\end{equation*}

A linear regression on the variables log(number of sites) and log(number of
pages) yielded [1.647, 1.853] as the 95\% confidence interval for the value
\ss\ in the Alexa crawl of 250,000 sites of the World Wide Web. For the
Infoseek crawl, the 95\% confidence interval for \ss\ is [1.775, 1.909].
These estimates for the value of $\beta $\ are consistent across the two
data sets and with the model, which predicts a linear dependence between the
logarithm of the variables to be linear with slope $-\beta <-1$.

\bigskip The existence of this universal power law has practical
consequences as well, since one can estimate the expected number of sites of
any arbitrary size, even if a site of that size has not yet been observed.
This can be achieved by extrapolating the power law given by Eq. 9, to any
large $n_{s}$, e.g. $P(n_{s2})=P(n_{s1})(n_{s1}/n_{s2})^{-\beta }$. The
expected number of sites of size $n_{s2}$ in a crawl of $N$ sites would be $%
NP(n_{s2})$. As an example, from the Alexa data we can infer that if one
were to collect data on 250,000 sites the probability of finding a site with
a million pages would be $10^{-4}$. Notice that this information is not
readily available from the crawl alone, since it stops at 105 pages per site.

Several points are worth making. First, since small values of $n_{s}$ lie
outside the scaling regime, our theory does not explain the data on sites
with few pages. Secondly, as a consequence of the universality of our
prediction, as more sites will be created, the same power law behavior will
be seen. This will once again allow for the determination of largest sites
from data that will be limited in scope due to server limitations. Finally,
since the process of ranking random variables stemming from any broad
distribution always produces a narrow and monotically decreasing power law
of the type originally discussed by Zipf (10), we expect that such ranking
will lead to a Zipf-like law(11).

In summary, we presented a stochastic theory of the growth dynamics of the
Web that takes into account the wide range of stochastic growth rates in the
number of pages per site, as well as the fact that new sites are created at
different times in the unfolding story of the Web. This leads to the
prediction of a universal power law in the distribution of the number of
pages per site, which we confirm experimentally by analyzing data from two
large crawls by the search engines Alexa and Infoseek. The existence of this
power law not only implies the lack of any length scale for the Web, but
also allows to estimate the number of sites of any given size without having
to exhaustively crawl the Web. This is yet a another example of the strong
regularities(12) that are revealed in studies of the Web, and which become
apparent because of its sheer size and reach. 
\begin{equation*}
\end{equation*}

\begin{acknowledgement}
We thank Jim Pitkow and Eytan Adar for providing data for our analysis, and
Rajan Lukose for many useful discussions.
\end{acknowledgement}

\ \ \qquad\ \ \ \ \

\end{document}